# Project X: A Flexible High Power Proton Facility
## The Project X Collaboration

Corresponding Author: Steve Holmes, holmes@fnal.gov

Project X is a high intensity proton facility that will support a world-leading Intensity Frontier research program over the next several decades at Fermilab. When compared to other facilities in the planning stages elsewhere in the world Project X is completely unique in its ability to deliver, simultaneously, up to 6 MW of site-wide beam power to multiple experiments, at multiple energies, and with flexible beam formats. Project X will support a wide range of experiments based on neutrinos, muons, kaons, nucleons, and nuclei. In addition, Project X will lay the foundation for the long-term development of a Neutrino Factory and/or Muon Collider. A complete concept for Project X has been developed and is documented in the Project X Reference Design Report (Project X RDR). The 2013 HEPAP Facilities Subpanel has assessed the science capabilities of Project X as "absolutely central" and the state of development as "ready for construction".

**Science:** Project X is an integral part of the U.S. Intensity Frontier Roadmap as developed through the P5 (P5 Report) and Fermilab Strategic Planning (Fermilab Plan for Discovery) processes. The primary mission elements associated with Project X include:

*Neutrino Experiments*: A high-power source of protons with energies between 1 and 120 GeV will produce intense neutrino beams illuminating both near and far detectors.
    *Goal*: At least 2 MW of proton beam power at any energy between 60 to 120 GeV; several hundred kW of proton beam power at 8 GeV.

*Kaon, Muon, Nuclei, and Nucleon Precision Experiments*: High power proton beams at 1 and 3 GeV will enable world-leading experiments pursuing ultra-rare muon and kaon decays, atomic, proton and neutron electron dipole moments (edms), and high sensitivity searches for neutron-antineutron oscillations.
    *Goal*: Multi-MW proton beams at 1 and 3 GeV, with flexible capability for providing distinct beam formats to multiple concurrent users. Simultaneous operation with the neutrino program is required.

*Platform for Evolution to a Neutrino Factory and Muon Collider*: A multi-MW proton facility is an integral part of the front end for a Neutrino Factory and/or Muon Collider.
    *Goal*: Provide a straightforward upgrade path for a 4 MW, low duty factor, source of protons at energies between 5 to 15 GeV.

*Material Science and Nuclear Energy Applications*: Accelerator, spallation target, and transmutation technology demonstrations are critical input into the design of future energy systems, including fission and fusion reactors, and nuclear waste transmutation systems.

Muon spin rotation techniques (muSR) can provide sensitive probes of the magnetic structure of materials.

Goal: Provide MW-class proton beams at 1 GeV, coupled with novel target station designs, to support a range of materials science and energy applications.

**Performance Capability:** The Project X Reference Design supports the mission elements and goals described above. The capabilities for the various programs are summarized in Table 1. An important feature of Project X is that these performance levels are simultaneously achievable.

**Table 1: Performance Goals for Project X**

| Main Injector Fast Spill (Long Baseline Neutrino Program) | | |
|---|---|---|
| Beam Energy | 60-120 | GeV |
| Beam Power | 2450 | kW |
| Protons per pulse | $1.5 \times 10^{14}$ | |
| Pulse length | 9.5 | μsec |
| Number of bunches | 504 | |
| Bunch spacing | 18.9 | nsec |
| Bunch length (FWHM) | 2 | nsec |
| Pulse repetition period | 0.6-1.2 | sec |
| **8 GeV Program (Short Baseline Neutrino Program)** | | |
| Beam Energy | 8 | GeV |
| Beam Power* | 0-172 | kW |
| Protons per pulse | $2.7 \times 10^{13}$ | |
| Pulse length | 4.3 | msec |
| Number of bunches | 140,000 | |
| Bunch spacing | 30 | nsec |
| Bunch length (FWHM) | 0.04 | nsec |
| Pulse repetition rate | 10 | Hz |
| **3 GeV Program (Muons, Kaons)** | | |
| Beam Energy | 3 | GeV |
| Beam Power | 2870 | kW |
| Protons per second | $6.2 \times 10^{15}$ | |
| Pulse length | CW | |
| Bunch spacing** | Programmable | |
| Bunch length (FWHM) | .04 | nsec |

| 1 GeV Program (neutrons, nucleons, materials) | |
|---|---|
| Beam Energy | 1 GeV |
| Beam Power | 1000 kW |
| Protons per second | $6.2 \times 10^{15}$ |
| Pulse length | CW |
| Bunch spacing** | Programmable |
| Bunch length (FWHM) | .04 nsec |

*Beam power available at 8 GeV is dependent on the operational energy of the Main Injector.

** Independent bunch patterns can be provided from the 1 and 3 GeV linacs to three experimental areas simultaneously

**Staging:** Budgetary constraints have led to development of a staged approach to Project X, based on application of the following principles:
- Each stage must present compelling physics opportunities;
- Each stage should be constructible for significantly less than $1B;
- Each stage should utilize existing elements of the Fermilab complex to the extent possible;
- Each stage should be constructible with minor interruptions to ongoing program;
- At the completion of the final stage the full Reference Design should be realized.

A three stage implementation of the Reference Design consistent with the above principles has been developed with significant scientific potential associated with all mission elements at each stage (see for example [Physics Opportunities with Stage 1 of Project X)](). The expansion of Table 1 to all three stages is given in [Project X Performance by Stage]().

The siting and configuration associated with the Project X Reference Design is strongly influenced by the staging strategy and a particular siting has been identified that maintains consistency with this plan. The siting is shown in Figure 1. As displayed in the figure the 1 GeV, 1-3 GeV, and 3-8 GeV linacs are physically distinct.

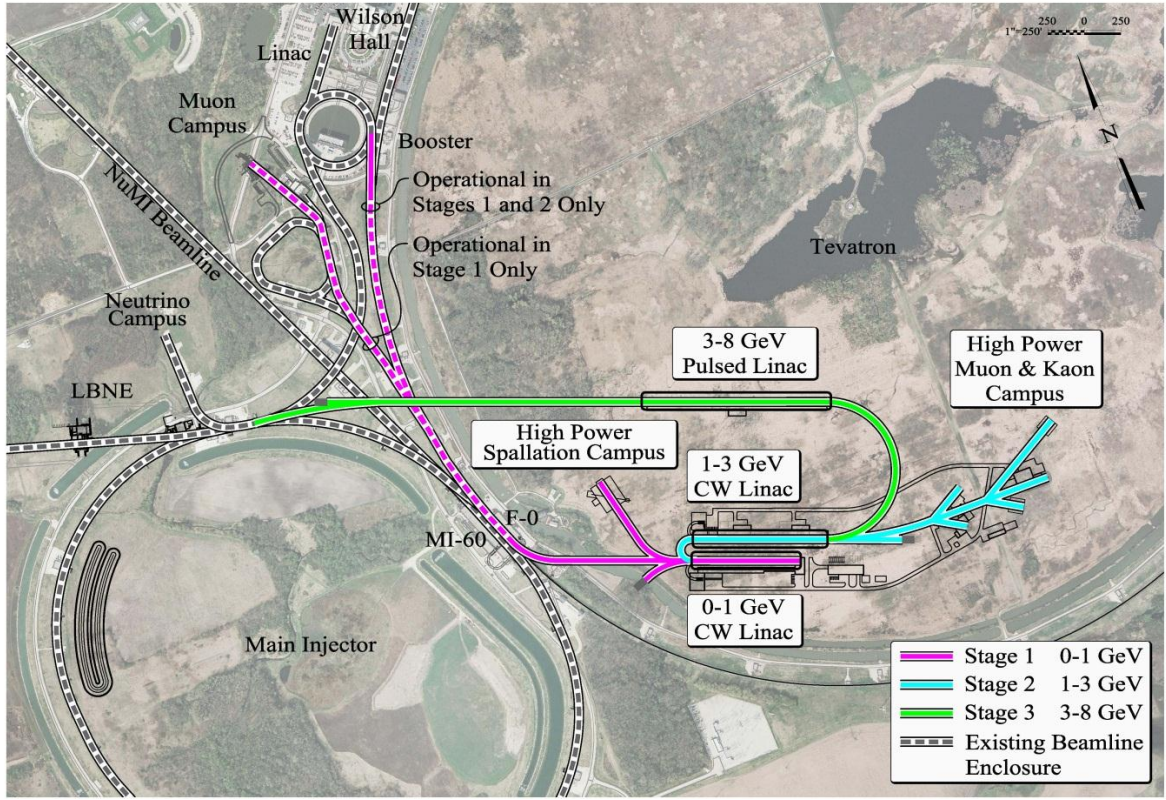

**Figure 1: Site layout for Project X**

**Development Program:** Project X capitalizes on the very rapid development of superconducting rf technologies over the last 20 years, and their highly successful application to high power H⁻ acceleration at the Spallation Neutron Source at Oak Ridge National Laboratory. As a result of these developments excellent simulation and modeling tools exist for designing the Project X facility with high confidence that performance goals can be achieved, and the primary supporting technologies required to construct Project X exist today.

The Project X Collaboration is engaged in a comprehensive development program aimed at mitigating technical and cost risks associated with construction and operations. The Reference Design provides the context for the R&D program; the primary elements of this program are:

- Accelerator Configuration and Performance Projections
    A complete beam-based design of the Project X facility has been developed including comprehensive electromagnetic modeling of components, and modeling and simulations of beam transmission through the complex. The configuration established through this program provides the context for all sub-system component development.

- Front End (0-25 MeV)

    The unique capabilities of Project X are derived from the performance of the Front End. A program of individual and integrated systems testing of front end components has been initiated (PXIE – Project X Injector Experiment).

- $H^-$ Injection

    $H^-$ injection into the first circular ring of Project X (the Booster in Stages 1 and 2, the Recycler in Stage 3) represents a particular challenge. Multiple concepts are being developed and tested, both through simulations and hardware development.

- High Intensity Recycler/Main Injector Operations

    The Recycler/Main Injector complex will be required to accelerate a factor of three more beam in the Project X era than current operations. Issues such as rf system requirements, space-charge, electron cloud, and a variety of potential beam instabilities are being investigated.

- High Power Targets

    Project X requires the development of targets capable of handling MWs of beam power. The development program has evolved from the LBNE development program, and includes an international consortium (RaDIATE – Radiation Damage In Accelerator Target Environments) and direct contributions from collaborators knowledgeable in the design and construction of spallation targets.

- Superconducting rf

    Six different accelerating structures operating at four different frequencies are required for Project X. A comprehensive program, originally initiated under ILC and undertaken with national and international partners, has been underway for a number of years. To date this program has produced single and multi-cell cavities meeting Project X specifications at 325 MHz, 650 MHz, and 1.3 GHz.

The overall scope and goals of the Project X development program are based on being prepared for a 2017 construction start. Essentially all elements listed above are required for Stage 1 implementation.